\documentclass[aps,twocolumn,raggedbottom,showpacs,nobalancelastpage,amssymb,groupaddress]{revtex4}

\usepackage[english]{babel}
\usepackage{graphicx}
\usepackage{amssymb}
\usepackage{amsmath}
\usepackage{amscd}
\usepackage{eucal}
\usepackage{color}
\usepackage{bm}

\usepackage{natbib}
\usepackage[bookmarks=true,colorlinks,linkcolor=red,urlcolor=blue,citecolor=blue]{hyperref}
\usepackage{dcolumn}

\def\be{\begin{equation}}
\def\ee{\end{equation}}
\def\bea{\begin{eqnarray}}
\def\eea{\end{eqnarray}}
\def\bsub{\begin{subequations}}
\def\esub{\end{subequations}}

\usepackage{ulem}

\begin{document}

\title{Competing topological phases in few-layer graphene}

\newcommand{\spsms}{CEA-INAC/UJF Grenoble 1, SPSMS UMR-E 9001, Grenoble F-38054, France}
\newcommand{\spec}{SPEC, CNRS URA 2464, IRAMIS-CEA Saclay, 91191 Gif-sur-Yvette, France}

\author{Pierre Carmier}
\author{Oleksii Shevtsov}
\author{Christoph Groth}
\author{Xavier Waintal}
\affiliation{\spsms}
\date{\today}

\begin{abstract}
We investigate the effect of spin-orbit coupling on the band structure of graphene-based two-dimensional Dirac fermion gases in the quantum Hall regime. Taking monolayer graphene as our first candidate, we show that a quantum phase transition between two distinct topological states -- the quantum Hall and the quantum spin Hall phases -- can be driven by simply tuning the Fermi level with a gate voltage. This transition is characterized by the existence of a chiral spin-polarized edge state propagating along the interface separating the two topological phases. We then apply our analysis to the more difficult case of bilayer graphene. Unlike in monolayer graphene, spin-orbit coupling by itself has indeed been predicted to be unsuccessful in driving bilayer graphene into a topological phase, due to the existence of an even number of pairs of spin-polarized edge states. While we show that this remains the case in the quantum Hall regime, we point out that by additionally breaking the layer inversion symmetry, a non-trivial quantum spin Hall phase can re-emerge in bilayer graphene at low energy. We consider two different symmetry-breaking mechanisms: inducing spin-orbit coupling only in the upper layer, and applying a perpendicular electric field. In both cases, the presence at low energy of an odd number of pairs of edge states can be driven by an exchange field. The related situation in trilayer graphene is also discussed.
\end{abstract}

\pacs{72.80.Vp, 73.43.Nq, 73.20.At, 73.21.Ac}

\maketitle

\section{Introduction}

Topological insulators are bulk insulators which possess robust conducting surface states \cite{Hasan10,Qi11}. Paradigmatic two-dimensional examples of this class are the quantum Hall (QH) and the quantum spin Hall (QSH) phases, which are characterized by respectively chiral and helical one-dimensional edge states. While the former can be generated by simply applying a strong perpendicular magnetic field and has been rather extensively studied since the 1980s, the latter requires the presence of spin-orbit coupling and has received very little experimental evidence. Indeed, despite the wide interest shown in the literature for the QSH phase (and for topological phases in general) since the seminal works by Kane and Mele \cite{Kane05bis,Kane05}, experimental traces of this phase have remained scarce, with the exception of the remarkable works on HgTe quantum wells \cite{Bernevig06,Konig07,Roth09} (see also the experiment involving InAs \cite{Knez11}). Recent studies \cite{Weeks11,Shevtsov12,Jiang12} have revived the possibility of generating a QSH in graphene \cite{CastroNeto09,DasSarma11}, by showing that low concentrations of suitably chosen adatoms, randomly deposited on graphene, could open a large non-trivial gap in graphene's otherwise semimetallic band structure, and yield transport properties showing no trace of the spatially inhomogenous spin-orbit coupling. The perspective of successfully turning graphene into a QSH insulator is promising, as it would considerably enhance the experimental feasibility of engineering samples of the latter, which are so far limited to the previously mentioned and experimentally challenging HgTe heterostructures.

Although it does not enjoy the conceptual simplicity of the monolayer, bilayer graphene is an interesting system in its own right. It is a gapless semimetal, characterized by massive chiral excitations carrying a topological Berry phase 2$\pi$ \cite{Novoselov06}, with a very rich list of many-body instabilities predicted at low density (see \cite{McCann12} for corresponding references). One of its most remarkable properties, in contrast to monolayer graphene, is the possibility to open a gap in its band structure by simply applying a perpendicular electric field which breaks the layer inversion symmetry \cite{Castro07}. However, in the presence of a perpendicular magnetic field, the electronic properties of both systems become qualitatively very similar. In particular, their energy spectrum is characterized by a particle-hole symmetry and by the existence of levels sitting exactly at zero energy which are at the origin of the anomalous quantization of the Hall conductance in these systems as compared to other two-dimensional electron gases \cite{Novoselov05,Zhang05,Gorbig11}. This is the hallmark property of what we will refer to in this article as two-dimensional Dirac fermion gases (2DDFGs).

The purpose of this article is to discuss what happens to the band structure of a 2DDFG when the effects of both magnetic field and spin-orbit coupling are taken into account simultaneously. Taking graphene as our first example, we shall review in section \ref{sec:MG} the results already published elsewhere \cite{Shevtsov12X}, according to which a topological phase transition takes place at low energy and can be tuned by simply varying the chemical potential. Then, in the following section, we shall investigate the related situation in bilayer graphene and show that the results obtained for the monolayer do not extend to it. This can be traced back to the fact that the topological invariant characterizing the QSH phase is non-trivial only if there are an odd number of pairs of edge states, which translates in multi-layer graphene into the condition of having an odd number of layers. Nevertheless, we will show in section \ref{sec:BandAsym} that by adding additional ingredients to our model, namely by breaking the layer inversion symmetry, a non-trivial QSH phase can be generated in bilayer graphene along with a corresponding topological phase transition. We stress that (almost) all of our results can be understood by simply looking at the band structures of the systems we study. Finally, section \ref{sec:Disc} discusses the possible extension of our approach to other systems, and we conclude in section \ref{sec:Conc}.

\section{Graphene}
\label{sec:MG}

Recent investigations of the interplay between QH and QSH phases in some specific examples of 2DDFGs \cite{Tkachov10,DeMartino11,Goldman11,Shevtsov12X} have led to surprising results. In these works, it was shown that the QSH phase can survive the presence of a perpendicular magnetic field and that the $\mathbb{Z}_2$ topological invariant \cite{Kane05} remains non-trivial for energies below the spin-orbit induced gap, despite the breaking of time-reversal symmetry\,\footnote{The persistence of the QSH phase in a different time-reversal-symmetry-breaking context was also observed \cite{Yang11}.}. As we will exemplify in the case of graphene, the origin of this intriguing result actually stems from the existence of zero-energy Landau levels: as soon as the spin-degeneracy of these levels is lifted, spin-polarized edge states characteristic of a QSH phase emerge\,\footnote{This observation was first made by Abanin et al. \cite{Abanin06} who realized that the mechanism of Zeeman splitting could fulfill this condition; unfortunately, the weakness of Zeeman splitting in graphene rendered the effect too small to be observable at low-to-moderate magnetic fields, while it is washed out by many-body effects at large magnetic fields.}.

\subsection{Model}
\label{sec:Mod}

\begin{figure}[]
\begin{center}
\includegraphics[angle=0,width=0.75\linewidth]{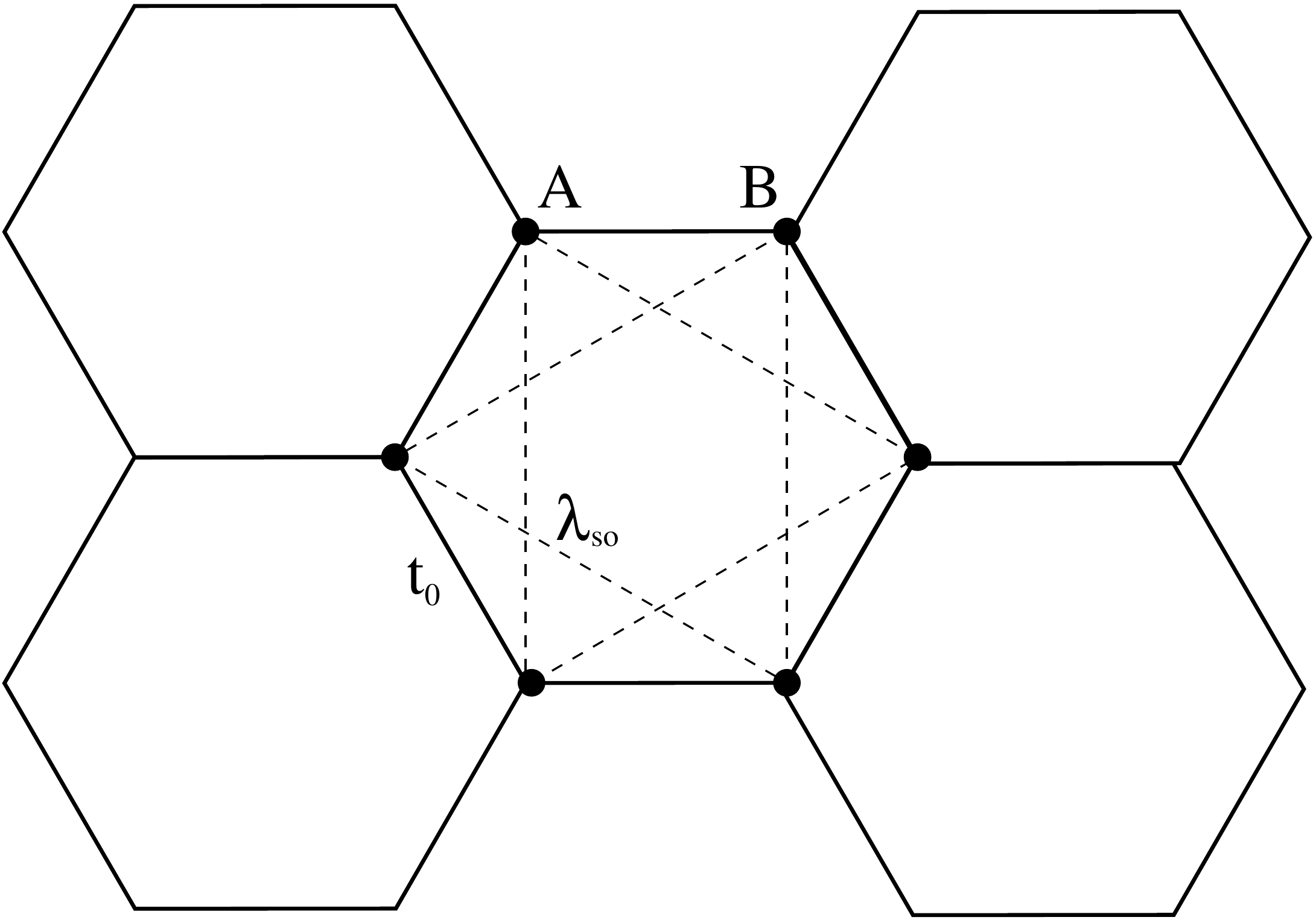}
\caption{Sketch of the graphene hexagonal lattice, characterized by nearest-neighbor hopping $t_0$ and next-nearest-neigbor Kane-Mele spin-orbit coupling $\lambda_{\text{so}}$. The unit cell of the lattice contains two inequivalent sites, A and B.}
\label{FigLatGra}
\end{center}
\end{figure}

Let us start by introducing the model from which our results shall be derived. In the vicinity of the zero-energy points in the Brillouin zone, low-energy excitations can be described by a Dirac Hamiltonian:
\be
\label{eq:MG}
H_{\text{G}} = v_F(\tau\hat{p}_x\sigma_x+\hat{p}_y\sigma_y) \; ,
\ee
with $\sigma_x,\sigma_y$ the usual set of Pauli matrices acting in the two-dimensional space of the inequivalent sublattices A and B (see Fig.~\ref{FigLatGra}), and with $v_F=\sqrt{3}t_0\tilde{a}/(2\hbar)$ the Fermi velocity, expressed as a function of the microscopic lattice parameters $t_0$ (nearest-neighbor hopping amplitude) and $\tilde{a}$ (lattice constant) which we choose in the following as our working units of energy and length, respectively. 
$\tau=\pm 1$ accounts for the two possible valleys from which the low-energy excitations can arise. In the vicinity of these points, the energy-dispersion relation reads $\epsilon = \pm v_F|{\bf p}|$.

The presence of a perpendicular magnetic field can be straightforwardly included by making use of the Peierls substitution $\hat{{\bf p}} \rightarrow \hat{{\bf \Pi}}=\hat{{\bf p}}+e{\bf A}$, which accounts for the presence of the magnetic vector potential ${\bf A}$ such that $\nabla\times{\bf A}=B{\bf z}$. The components of the generalized momentum satisfy the Heisenberg algebra $[\hat{\Pi}_x, \hat{\Pi}_y] = -i(\hbar/l_B)^2$. By expressing these components in terms of the usual harmonic oscillator ladder operators \cite{Gorbig11}, 
\be
\hat{\Pi}_x = \frac{\hbar}{\sqrt{2}l_B}(a+a^\dagger) \; , \;
\hat{\Pi}_y = \frac{i\hbar}{\sqrt{2}l_B}(a-a^\dagger) \; , 
\ee
and using the standard raising and lowering properties of these operators on the eigenstates ($a|n\rangle=\sqrt{n}|n-1\rangle$ and $a^\dagger|n-1\rangle=\sqrt{n}|n\rangle$), the energy spectrum can then straightforwardly be shown to turn into the well-known Landau levels, 
\be
\label{eq:LLMG}
\epsilon_n = \pm\Delta_B\sqrt{|n|}
\ee 
with $\Delta_B = \sqrt{2}\hbar v_F/l_B$, and $l_B=\sqrt{\hbar/(eB)}$ the magnetic length. As already mentioned before, the main distinctive feature of the Landau level spectrum of a 2DDFG as compared to that of a standard two-dimensional electron gas is the existence of a zero-energy level at $n=0$ originating from the pseudo-relativistic nature of the charge carriers. Also note that all levels enjoy a 4-fold degeneracy arising from spin and valley indices.

\subsection{Band structure / edge state correspondence}

The appearance of edge states in this context can be best understood by looking at the band structure of a graphene ribbon. The latter is a system which is translationally invariant in one direction, and confined in the other. In order to derive the band structure numerically, we formulate the above ingredients in terms of a tight-binding model, in which Eq.~(\ref{eq:MG}) becomes
\be
\label{eq:tbMG}
{\cal H}_{\text{G}} = -t_0\sum_{\langle i, j \rangle} e^{i\phi_{ij}} c_i^\dagger c_j \; .
\ee
Indices ($i,j$) label lattice sites, while symbol $\langle\;\rangle$ refers to nearest-neighbor coupling (with hopping amplitude $t_0$), as is illustrated in Fig.~\ref{FigLatGra}. The Peierls phase $\phi_{ij}=(e/\hbar)\int_{{\bf r}_j}^{{\bf r}_i}{\bf A}\cdot d{\bf r}$ takes into account the contribution from the magnetic flux threading the lattice. Numerical calculations are performed using kwant, the new quantum transport software package developed by A. Akhmerov, C. Groth, X. Waintal, and M. Wimmer. In the process, we choose to work with armchair boundary conditions, but our results are qualitatively unaffected by this choice.

\begin{figure}[]
\begin{center}
\includegraphics[angle=0,width=1.0\linewidth]{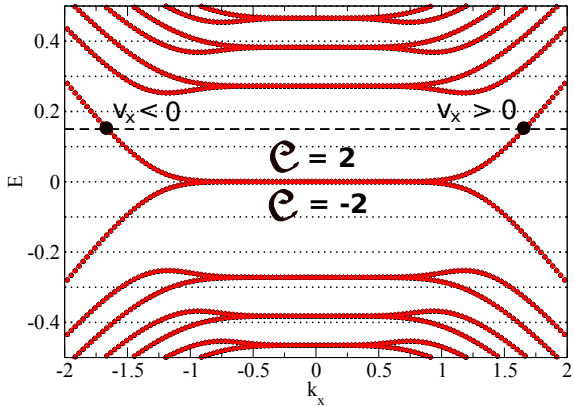}
\caption{(Color online): Energy spectrum of a monolayer graphene armchair ribbon in the QH regime (${\cal C}\neq0$). Black circles and red triangles respectively stand for spin up and spin down bands, which are here indistinguishable due to spin degeneracy. The horizontal dashed line represents an arbitrarily chosen Fermi level. Its intersection with the lowest band of the system indicates the existence of a (spin-degenerate) edge state which propagates in opposite directions on opposite sides. The ribbon width is $W=38$ and the magnetic length is $l_B\approx4$.}
\label{Fig1}
\end{center}
\end{figure}

The band structure associated with the peculiar spectrum of Eq.~(\ref{eq:LLMG}) in a ribbon geometry (translationally invariant in the $x$-direction, confined in the $y$-direction with $|y|<W/2$) is shown in Fig.~\ref{Fig1}. 
Due to the nature of the classical dynamics, the transverse coordinate of the cyclotronic center of motion $y_c$ can be identified with the conserved longitudinal momentum $k_x$ via the formula $y_c = -k_xl_B^2$. Observing the band structure in terms of this real space coordinate, one can see in Fig.~\ref{Fig1} that, far from the edges of the ribbon, the band structure consists of flat bands which are none other than the Landau levels of Eq.~(\ref{eq:LLMG}): electrons in the bulk are classically localized by the magnetic field along closed cyclotronic orbits. On the other hand, electrons in the vicinity of the edges can scatter along them and propagate following skipping orbits, which translates in the band structure into bulk Landau levels acquiring a finite dispersion as they approach the edges of the ribbon:
\be
\label{eq:disp}
v_x^{(n)} = \frac{1}{\hbar}\frac{\partial \epsilon_n}{\partial k_x} \; .
\ee 
Because this dispersion is monotonous on a given edge (see Fig.~\ref{Fig1}), the edge states cannot be backscattered unless they are coupled to the states living on the opposite edge, a process the likelihood of which decays exponentially with the width of the system. This property of the edge states is generally referred to as chirality and is the reason why these states can carry current without dissipation: this leads to the celebrated QH effect \cite{Klitzing80}, characterized by a quantized conductance $G={\cal C}(e^2/h)$. More formally, the edge states enjoy a topological protection encoded in the Chern number ${\cal C}$ which is a $\mathbb{Z}$ topological invariant characterizing the number of filled bands in the QH regime \cite{Thouless82}. It is a topological quantity, in the sense that smooth deformations of the Hamiltonian (deformations which do not close the gap) cannot change its value, and shall be defined in the next subsection. We thus see that, for most purposes, the physics of topological phases such as the QH phase can be very simply extracted from the corresponding band structure. 

\begin{figure}[]
\begin{center}
\includegraphics[angle=0,width=1.0\linewidth]{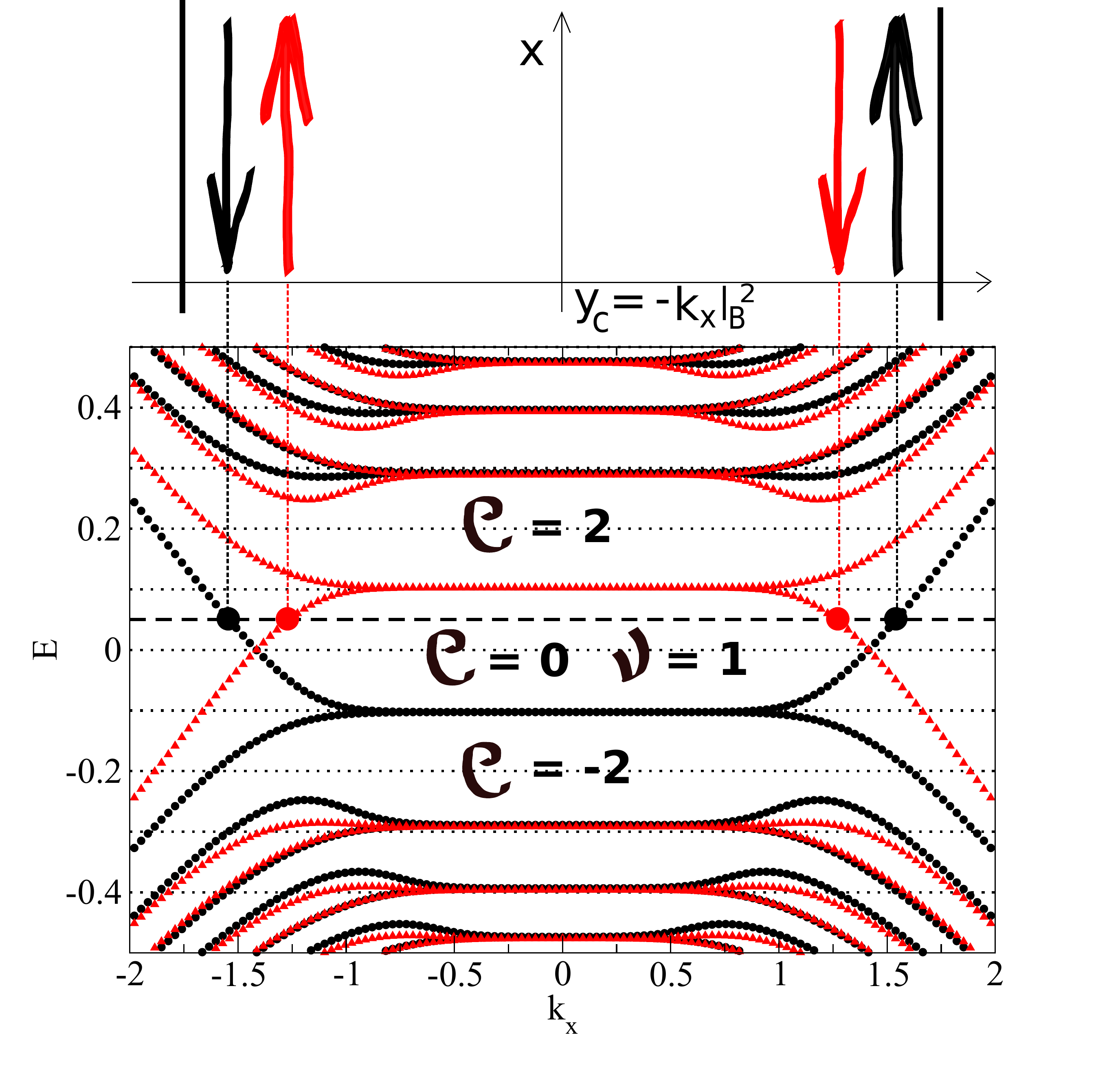}
\caption{(Color online): Same as in Fig.~\ref{Fig1}, but with an additional spin-orbit coupling term $\lambda_{\text{so}}=0.02$. The latter lifts the spin-degeneracy of the zero-energy Landau level, yielding a QSH phase ($\nu=1$) with a single pair of counter-propagating spin-polarized edge states. The sketch above the band structure depicts the edge states in real-space (thick black vertical lines on the sides represent the edges of the ribbon).}
\label{Fig1bot}
\end{center}
\end{figure}

We now consider the situation where, in addition to the perpendicular magnetic field, the effect of spin-orbit coupling as introduced by Kane and Mele \cite{Kane05bis} is accounted for in the Hamiltonian as
\be
\label{eq:KM0}
H_{\text{so}} = \tau s\Delta_{\text{so}}\sigma_z \; ,
\ee
which is characterized by the energy scale $\Delta_{\text{so}}$. $\tau=\pm 1$ and $s=\pm 1$ account for valley and spin degrees of freedom. In terms of a tight-binding model, Eq.~(\ref{eq:KM0}) can be implemented as \cite{Kane05bis}
\be
\label{eq:tbKM0}
{\cal H}_{\text{so}} = i\lambda_{\text{so}}\sum_{ \langle\langle i,j\rangle\rangle}\nu_{ij}e^{i\phi_{ij}} (c_{i,\alpha}^\dagger s^{\alpha\beta}_z c_{j,\beta}) \; ,
\ee
where indices ($i,j$) once more label lattice sites, while ($\alpha,\beta$) label spin indices, symbol $\langle\langle\;\rangle\rangle$ refers to next-nearest-neighbor coupling (with SO-induced hopping amplitude $\lambda_{\text{so}} = \Delta_{\text{so}}/(3\sqrt{3})$ \cite{Kane05bis}), and $\nu_{ij}=\pm1$ depending on whether sites are coupled clockwise or counter-clockwise (see Fig.~\ref{FigLatGra}). Note that in order for the system to remain gauge invariant, Peierls substitution has to be done on all hopping matrix elements: nearest-neighbor {\it and} (SO) second nearest-neighbor. The presence of spin-orbit coupling modifies the Landau level spectrum according to the expression 
\be
\epsilon_{n,s} = 
\left\lbrace
\begin{split}
\pm\sqrt{\Delta_B^2|n| + \Delta_{\text{so}}^2} \; , \; \text{for} \; n\neq0
\\
-s\Delta_{\text{so}} \; , \; \text{for} \; n=0
\end{split}
\right. \; .
\ee
The latter is characterized by the $n=0$ level being lifted from zero energy into spin-polarized branches: $E=+\Delta_{\text{so}}$ features only spin-down states, while $E=-\Delta_{\text{so}}$ features only spin-up states \cite{DeMartino11} (see Fig.~\ref{Fig1bot}). 
While other Landau levels retain their associated chiral edge states, irrespective of the spin polarization, the lowest Landau level now features counter-propagating edge states for $|E|< \Delta_{\text{so}}$, with a spin-dependent direction of propagation (see Fig.~\ref{Fig1bot}): one has $v_x^{(0,\uparrow)}\cdot v_x^{(0,\downarrow)} <0$ on a given edge (allowing for an additional spin dependence in the definition of Eq.~(\ref{eq:disp})). This is illustrated in the real-space sketch above the band structure in Fig.~\ref{Fig1bot}. It is the signature of a QSH phase, as can be certified by computing the associated $\mathbb{Z}_2$ topological invariant introduced by Kane and Mele \cite{Kane05}, which we do next.

\subsection{Topological order}

Let us start by recalling the standard topological number characterization of Landau levels when $\Delta_{\text{so}} = 0$. Each Landau level $n$ and its associated eigenfunctions over the first Brillouin zone are characterized by a topological invariant, the so-called Chern number \cite{Thouless82}. This topological number takes a value ${\cal C}^{(n)}_{\tau,s}=+1$ for each Landau level, independently of the Landau $n$, valley $\tau$ or spin $s$ indices. For each value of the Fermi energy, we can characterize the corresponding phase by a topological number 
\be
{\cal C} = \sum_{\tau,s} {\cal C}_{\tau,s} \; , \; \text{with} \; {\cal C}_{\tau,s}(E_F) = \sum_{\epsilon_n<E_F}{\cal C}^{(n)}_{\tau,s}
\ee
obtained by summing the Chern numbers of all filled energy bands \cite{Thouless82}.

In 2DDFGs, the Chern number is a priori ill-defined because of the existence of an infinite number of filled energy bands of negative energy. Through the use of non-commutative Berry's connection, it was however shown \cite{Watanabe11} that the Chern number takes a value ${\cal C}_{\tau,s}(E=0^-)=-1/2$ per degree of freedom for energies immediately below the Dirac point. In the case of graphene, defining the spin Chern number as
\be
{\cal C}_s=\sum_{\tau}{\cal C}_{\tau,s} \; ,
\ee
one obtains ${\cal C}_s(E=0^-)=-1$ per spin species (since there are two valleys). With this prescription, the band structure of graphene in the QH regime can be easily described by computing the value of the Chern number as a function of the Fermi energy. This yields 
\be
\left\lbrace
\begin{array}{l}
{\cal C}_\uparrow={\cal C}_\downarrow=-1 \; , \; \text{for} \; -\Delta_B < E_F < 0
\\
{\cal C}_\uparrow={\cal C}_\downarrow=2n+1 \; , \; \text{for} \; E_F > 0
\end{array}
\right.
\ee
where $n$ is the index of the highest filled Landau level. Notice that Chern numbers of each spin species are equal (since the spectrum is spin-degenerate) and that ${\cal C}={\cal C}_\uparrow+{\cal C}_\downarrow$ is always non-zero, as expected for a QH phase. The total Chern number increases step-wise by multiples of 4, due to spin and valley degeneracy of the Landau levels.

Restoring a finite value of $\Delta_{\text{so}}$, the topologically non-trivial nature of the phase for $E<\Delta_{\text{so}}$ can be checked by computing the corresponding value of the $\mathbb{Z}_2$ topological invariant. In the presence of spin rotational symmetry (conservation of $S_z$), this invariant can be simply expressed as the difference of the Chern numbers for each spin species \cite{Sheng06}: 
\be
\nu=\frac{1}{2}({\cal C}_\uparrow-{\cal C}_\downarrow) \;\; (\text{mod} \; 2) \; .
\ee
Note, however, that the existence of this invariant is naturally independent of the existence of this symmetry. Following the calculations performed in \cite{Shevtsov12X}, one easily finds
\be
\left\lbrace
\begin{array}{l}
{\cal C}_\uparrow=-{\cal C}_\downarrow=+1 \; , \; \text{for} \; |E_F| < \Delta_{\text{so}}
\\
{\cal C}_\uparrow={\cal C}_\downarrow=2n+1 \; , \; \text{for} \; E_F > \Delta_{\text{so}}
\end{array}
\right.
\ee
where $n$ is once more the index of the highest filled Landau level. This time, one is faced with a QH phase for $|E_F| > \Delta_{\text{so}}$, characterized by the same Chern number as in the previous case, while for $|E_F| < \Delta_{\text{so}}$, the total Chern number vanishes ${\cal C}_\uparrow+{\cal C}_\downarrow=0$, indicating that this region is no longer in the QH phase. However, as the $\mathbb{Z}_2$ invariant $\nu = 1$ does not vanish, the phase in this region is a QSH phase.

The resulting band structure is thus particularly interesting: it consists of a QSH phase at energies $|E|<\Delta_{\text{so}}$ and a QH phase at other energies. One can therefore observe a topological phase transition in this system by simply tuning the Fermi level across the spin-orbit gap. The central manifestation of this phase transition is the existence of a spin-polarized state localized at the interface between both phases. This state could be most clearly observed experimentally by making use of an additional electric gate to independently tune the Fermi level in two different parts of the system, thereby realizing a topological heterojunction. We refer the interested reader to our previous paper \cite{Shevtsov12X} for a detailed discussion of these matters.

\section{Bilayer graphene}
\label{sec:BandSym}

We now switch to the slightly more involved case of (Bernal-stacked) bilayer graphene and start by addressing the possibility of inducing a QSH phase in bilayer graphene. This is not a trivial endeavour, as the naive extension of the Kane-Mele model to bilayer graphene yields a weak $\mathbb{Z}_2$ topological phase, characterized by an even number (rather than an odd number, as in monolayer graphene) of pairs of spin-polarized edge states \cite{Prada11,Cortijo10}. This doubling of the number of edge states basically arises because a bilayer has twice as many layers as a monolayer. For the same reason, a graphene trilayer will have an odd number of pairs of edge states and therefore feature a non-trivial QSH phase. Breaking the layer symmetry by considering the case where spin-orbit coupling is present in only one of the layers was shown not to be any more effective \cite{Prada11}, the system then remaining semimetallic. 

Here we follow a different approach, inspired by the model presented in the previous section. This seems like a natural idea, as bilayer graphene is also known to feature zero-energy Landau levels \cite{Novoselov06,McCann06}. In this section, we will show that the presence of both spin-orbit coupling and a perpendicular magnetic field in the bilayer yields a band structure very similar to that of monolayer graphene, but with results no different from that of Prada et al. \cite{Prada11}: the obtained QSH phase is topologically trivial due to the existence of an even number of pairs of spin-polarized edge states. On the other hand, we will show in the next section that if spin-orbit coupling is present in only one of the two layers or if a perpendicular electric field is applied, then the breaking of layer inversion symmetry opens the door for a non-trivial QSH phase to arise at low-energy.

\subsection{Model}

\begin{figure}[]
\begin{center}
\includegraphics[angle=0,width=0.75\linewidth]{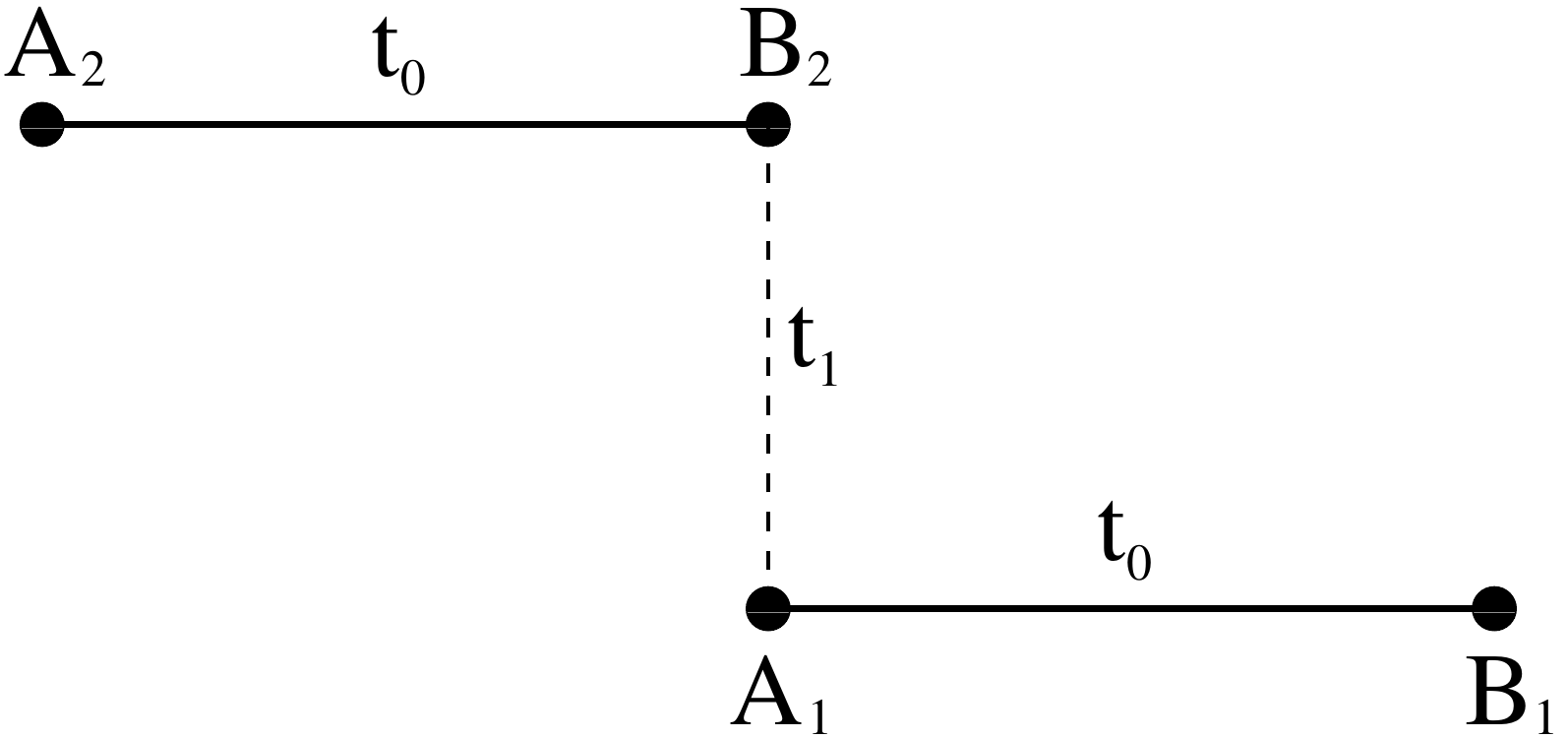}
\caption{Side view of Bernal-stacked bilayer graphene, which is characterized by intra-layer nearest-neighbor hopping $t_0$ and inter-layer hopping $t_1$.}
\label{FigBernal}
\end{center}
\end{figure}

The Hamiltonian for bilayer graphene can be expressed using two sets of Pauli matrices $\{\sigma,\eta\}$ which respectively refer to sublattice ($A$, $B$) and layer (1, 2) spaces. We consider the usual Bernal stacking, inherited from graphite (see Fig.~\ref{FigBernal}), in which $A$ atoms in the upper layer (2) lie above $B$ atoms from the lower layer (1).
Starting from the basis ($A_2$, $B_2$, $A_1$, $B_1$)$^T$, the Hamiltonian reads \cite{McCann06}:
\be
\label{eq:BG}
H_{\text{BG}} = v_F(\tau\hat{p}_x\sigma_x+\hat{p}_y\sigma_y)\eta_0 + \frac{t_1}{2}(\sigma_x\eta_x - \sigma_y\eta_y) \; .
\ee
The first term describes the usual low-energy Dirac structure of monolayer graphene. The second term takes into account the coupling between both layers, characterized by an energy scale $t_1 \simeq 0.15$. Corrections to Eq.~(\ref{eq:BG}) such as trigonal warping are small effects, typically only relevant below the meV range \cite{McCann12}, and will therefore be neglected.

The spectrum associated with Eq.~(\ref{eq:BG}) is particle-``hole" symmetric, with high-energy bands at $\epsilon^{\text{high}}=\pm t_1$ and low-energy bands touching at two Dirac points characterized by a topological Berry phase 2$\pi$. In the vicinity of this point, the energy-dispersion relation is quadratic, $\epsilon^{\text{low}} = \pm p^2/(2m^*)$, with $m^*=t_1/(2v_F^2)$ the effective mass of the gapless excitations. 

The presence of a perpendicular magnetic field can be straightforwardly included by making use of the Peierls substitution as before, and the energy spectrum can then be shown to turn into the well-known Landau levels \cite{McCann06}, 
\be
\label{eq:LLBG}
\epsilon_n = \pm\hbar\omega_c\sqrt{n(n-1)}
\ee 
with $\omega_c=eB/m^*$ the characteristic cyclotronic frequency. The latter can be related to the monolayer graphene energy scale by the simple relation $\hbar\omega_c=\Delta_B^2/t_1$. Notice how the spectrum in Eq.~(\ref{eq:LLBG}) features twice as many zero-energy levels as in monolayer graphene, since the $n=1$ level also vanishes. Actually, one can prove on general grounds that chirally stacked $N$-layer graphene should feature a $4N$-fold degenerate zero-energy Landau level \cite{Min08,Koshino09}. One should also have in mind that the accuracy of expression (\ref{eq:LLBG}) for $n\neq0, 1$ is only correct in the limit $\Delta_B \ll t_1$.

\subsection{Quantum Hall regime}

\begin{figure}[]
\begin{center}
\includegraphics[angle=0,width=1.0\linewidth]{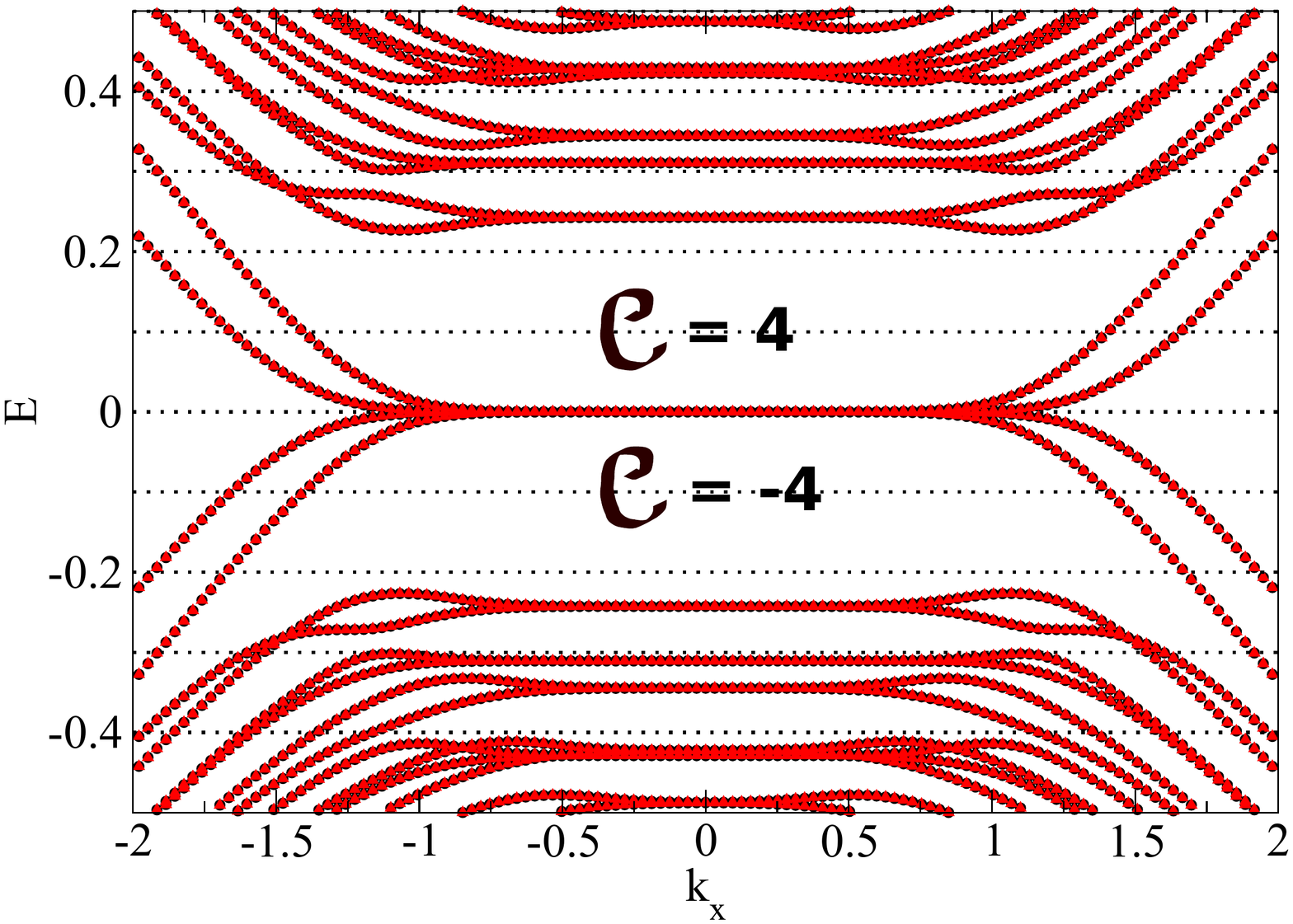}
\includegraphics[angle=0,width=1.0\linewidth]{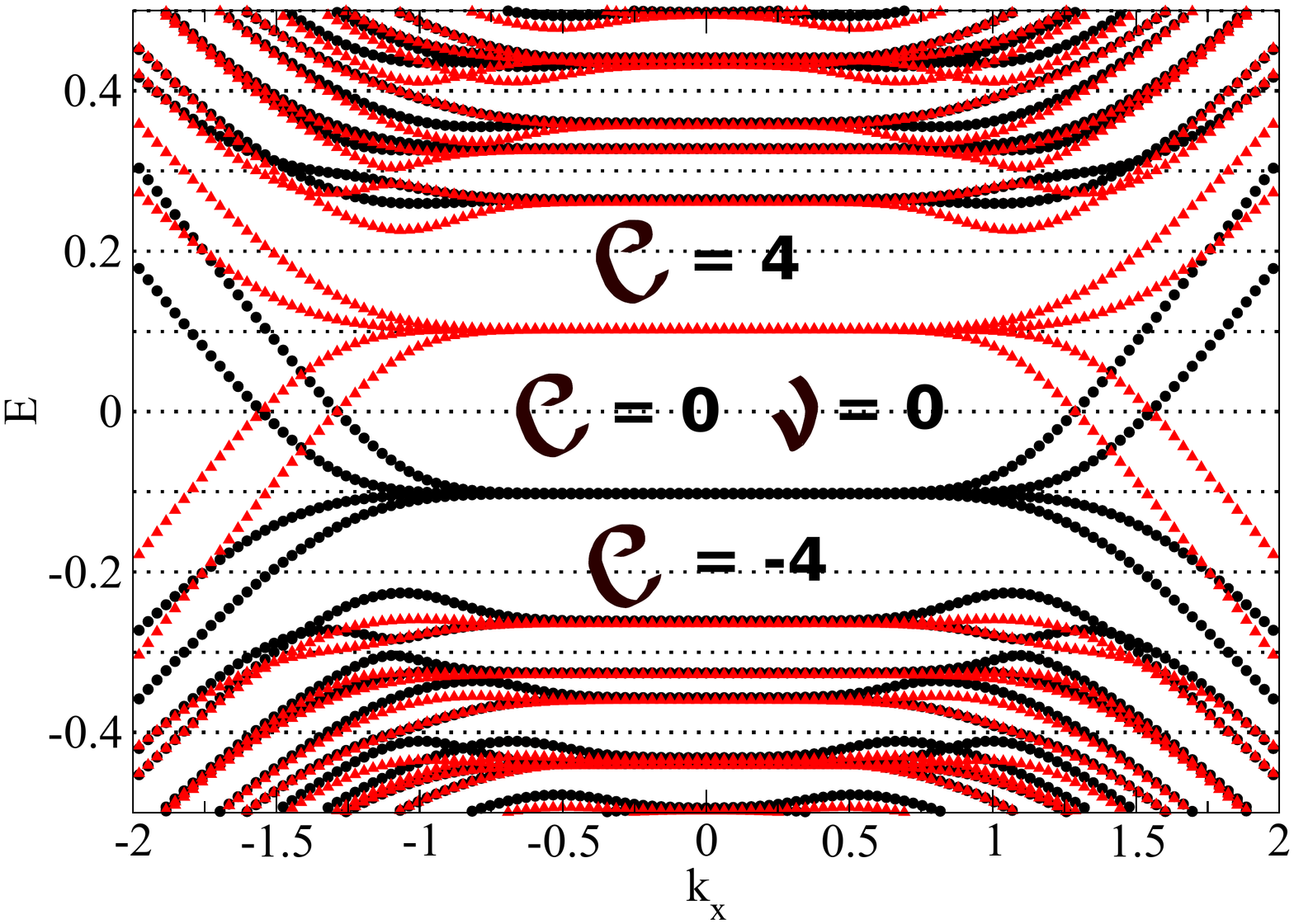}
\caption{(Color online): (Top panel) Energy spectrum of a bilayer graphene armchair ribbon in the QH regime. Black circles and red triangles respectively stand for spin up and spin down bands, which are here indistinguishable due to spin degeneracy. Notice that the zero-energy Landau level has twice as many bands as its counterpart in monolayer graphene. The ribbon width is $W=38$ and the magnetic length is $l_B\approx4$. (Bottom panel) Same as above, but with an additional spin-orbit coupling term $\lambda_{\text{so}}=0.02$. The latter lifts the spin-degeneracy of the zero-energy Landau level, yielding a weak QSH phase ($\nu=0$ (mod 2)) with an even number of pairs of counter-propagating spin-polarized edge states.}
\label{Fig2}
\end{center}
\end{figure}

To compute the associated band structure numerically, we once more formulate the above ingredients in terms of a tight-binding model, in which Eq.~(\ref{eq:BG}) becomes
\be
\label{eq:tbBG}
{\cal H}_{\text{BG}} = -t_0\sum_{\langle i, j \rangle} e^{i\phi_{ij}} c_i^\dagger c_j + t_1\sum_{\langle i\in A_2, j\in B_1 \rangle} c_i^\dagger c_j \; ,
\ee
using the same notations as in the previous section. No Peierls phase appears in the second term, as $A_2$-$B_1$ bonds are oriented along the $z$-axis. The band structure associated with the Landau level spectrum of Eq.~(\ref{eq:LLBG}) in a ribbon-geometry is displayed in the upper panel of Fig.~\ref{Fig2}. 
As expected, it resembles very closely that of monolayer graphene in the QH regime. The main difference between the two lies in the existence of twice as many dispersing branches arising from the lowest Landau level in bilayer graphene, due to the doubling of the zero-energy Landau level degeneracy. This translates in the language of topological invariants into Chern numbers per spin species ${\cal C}_s=-2$ immediately below zero energy. With this prescription, the band structure of bilayer graphene in the QH regime can be easily described by computing the value of the Chern number as a function of the Fermi energy, yielding: 
\be
\left\lbrace
\begin{array}{l}
{\cal C}_\uparrow={\cal C}_\downarrow=-2 \; , \; \text{for} \; -\hbar\omega_c\sqrt{2} < E_F < 0
\\
{\cal C}_\uparrow={\cal C}_\downarrow=2[\text{max}(0,n-1)+1] \; , \; \text{for} \; E_F > 0
\end{array}
\right.
\ee
where $n$ is the index of the highest filled Landau level. Notice that Chern numbers of each spin species are equal (since the spectrum is spin-degenerate) and that ${\cal C}={\cal C}_\uparrow+{\cal C}_\downarrow$ is always non-zero, as expected for a QH phase.

\subsection{Effect of layer-symmetric spin-orbit coupling}

We now consider the situation where, in addition to the perpendicular magnetic field, the effect of spin-orbit coupling as introduced by Kane and Mele for monolayer graphene is accounted for symmetrically in both layers\,\footnote{As in monolayer graphene, nearly identical results can be obtained by considering Zeeman splitting, instead of spin-orbit coupling, as the spin-degeneracy lifting mechanism.}. The layer-degenerate Kane-Mele spin-orbit coupling term is encoded in the Hamiltonian
\be
\label{eq:KM}
H_{\text{so}} = \tau s\Delta_{\text{so}}\sigma_z\eta_0 \; ,
\ee
which, in terms of a tight-binding model, can be implemented as
\be
\label{eq:tbKM}
{\cal H}_{\text{so}} = i\lambda_{\text{so}}\sum_{ \langle\langle i,j\rangle\rangle}\nu_{ij}e^{i\phi_{ij}} (c_{i,\alpha}^\dagger s^{\alpha\beta}_z c_{j,\beta}) \; ,
\ee
with similar notations as in the previous section, symbol $\langle\langle\;\rangle\rangle$ referring to intra-layer next-nearest-neighbor coupling. The presence of spin-orbit coupling modifies the Landau level spectrum according to the expression 
\be
\epsilon_{n,s} = 
\left\lbrace
\begin{split}
\pm\sqrt{n(n-1)(\hbar\omega_c)^2 + \Delta_{\text{so}}^2} \; , \; \text{for} \; n \neq 0, 1
\\
-s\Delta_{\text{so}} \; , \; \text{for} \; n = 0, 1
\end{split}
\right. \; .
\ee
The latter is characterized by $n=0$ and $n=1$ levels lifted from zero energy into spin-polarized branches: $E=+\Delta_{\text{so}}$ features only spin-down states, while $E=-\Delta_{\text{so}}$ features only spin-up states (see lower panel of Fig.~\ref{Fig2}).

The topologically trivial nature of the corresponding low-energy phase can be checked by computing the value of the $\mathbb{Z}_2$ topological invariant. As a straightforward generalization of the calculations performed in \cite{Shevtsov12X}, one obtains the following results:
\be
\left\lbrace
\begin{array}{l}
{\cal C}_\uparrow=-{\cal C}_\downarrow=+2 \; , \; \text{for} \; |E_F| < \Delta_{\text{so}}
\\
{\cal C}_\uparrow={\cal C}_\downarrow=2[\text{max}(0,n-1)+1] \; , \; \text{for} \; E_F > \Delta_{\text{so}}
\end{array}
\right.
\ee
where $n$ is once more the index of the highest filled Landau level. This time, one is faced with a QH phase for $|E_F| > \Delta_{\text{so}}$, characterized by the same Chern number as in the absence of spin-orbit coupling, while for $|E_F| < \Delta_{\text{so}}$, the total Chern number vanishes ${\cal C}_\uparrow+{\cal C}_\downarrow=0$, indicating that this region is no longer in the QH phase. However, as the $\mathbb{Z}_2$ invariant $\nu = 0$ (mod 2) also vanishes, the phase in this region is not a QSH phase either: rather, it is a weak QSH phase, in the sense that time-reversal-symmetric perturbations can couple the edge states and induce backscattering. This was not the case in monolayer graphene, due to the existence in the latter of a single pair of counter-propagating spin-polarized edge states at low energy. In the situation discussed in this section, we are thus led to conclude that a similar picture as that described in Ref.~\cite{Prada11} prevails. The way around this involves breaking the layer inversion symmetry, as we will see in the next section.

Before moving on, however, we would like to pause and comment on the fact that the model we considered in this section could provide a convenient platform for testing precisely how weak a $\nu=0$ (mod 2) QSH phase would be. Indeed, even though theory predicts that pairs of edge states should couple through backscattering processes, an experimental measure of how strongly edge state transport would be destroyed by such processes is yet to be done, and one cannot exclude the possibility of unexpected robustness, similarly to what has recently begun to be understood in so-called weak three-dimensional topological insulators \cite{Ringel12,Mong12}. Said a little differently, it remains unclear how one could distinguish through transport measurements a topological phase from a trivial phase which has edge states (such as the one exhibited in this section).

\section{Bilayer graphene with broken layer inversion symmetry}
\label{sec:BandAsym}

This section is devoted to the study of two layer inversion symmetry breaking mechanisms which enable an exchange-induced QSH phase to arise at low energy: (i) inducing spin-orbit coupling only in one of the layers, and (ii) applying a perpendicular electric field. We provide estimations of the magnitude of the QSH gap, and also address other possible phases which appear in our settings: a spin-polarized QH phase and a quantum valley Hall phase.

\subsection{Mechanism I: layer-asymmetric spin-orbit coupling}

Using the same basis as in Eq.~(\ref{eq:BG}), we now replace Eq.~(\ref{eq:KM}) by the symmetry-breaking term
\be
\label{eq:asymKM}
H_{\text{so}}^{\text{asym}} = \tau s\Delta_{\text{so}}\sigma_z\left(\frac{\eta_0+\eta_z}{2}\right)
\ee
which induces spin-orbit coupling only in the upper layer. This situation is physically relevant if one considers the possibility of inducing spin-orbit coupling in graphene by depositing adatoms on the surface \cite{Weeks11,Shevtsov12,Jiang12}. The corresponding tight-binding expression is given by applying Eq.~(\ref{eq:tbKM}) only in the upper layer.

\begin{figure}[]
\begin{center}
\includegraphics[angle=0,width=1.0\linewidth]{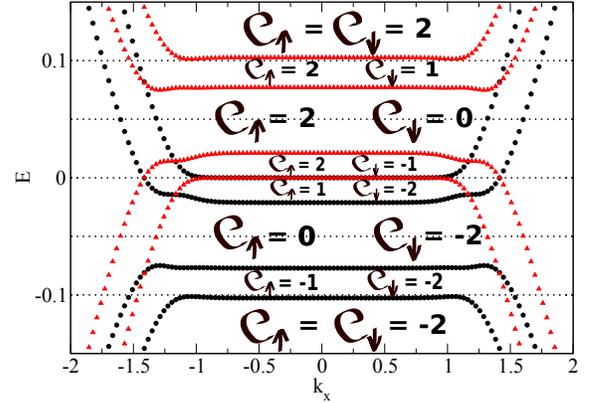}
\includegraphics[angle=0,width=1.0\linewidth]{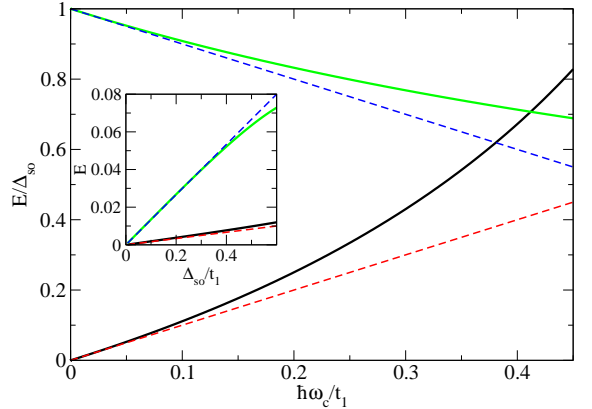}
\caption{(Color online):  (Top panel) Band structure with same parameters as in Fig.~\ref{Fig2}, except that spin-orbit coupling is only applied in the upper layer. This time, a spin-degenerate zero-energy Landau level survives, while low-energy edge states are characterized by an unbalanced spin population on a given edge ($|{\cal C}_\uparrow| \neq |{\cal C}_\downarrow|$). (Bottom panel) Dependence of the eigenvalues $\epsilon_{\pm}$ on the magnetic field for $\Delta_{\text{so}}=0.01$. Dashed lines correspond to the analytical predictions (valid in the perturbative limit $\hbar\omega_c \ll t_1$) and thick lines to the numerically obtained values. Inset: Dependence of $\epsilon_\pm$ on the spin-orbit gap for $\Delta_B=0.05$.}
\label{Fig3}
\end{center}
\end{figure}

In order to obtain the new Landau level spectrum, using the ladder operators introduced in section \ref{sec:MG}, one must now solve a quartic equation $\epsilon_n^4 + \alpha_n\epsilon_n^2 + \beta_n\epsilon_n + \gamma_n = 0$, with a non-vanishing linear term $\beta_n\neq0$:
\be
\left\lbrace
\begin{array}{l}
\alpha_n = -\left(t_1^2 + \Delta_{\text{so}}^2 + (2n-1)\Delta_B^2\right)
\vspace*{0.2cm}
\\
\beta_n = -s\Delta_{\text{so}}t_1^2
\vspace*{0.2cm}
\\
\gamma_n = (n-1)\Delta_B^2(n\Delta_B^2 + \Delta_{\text{so}}^2)
\end{array}
\right.
\; .
\ee
Note that the role of opposite spin polarizations is exchanged when going from the conduction to the valence band: $\epsilon_n(-s)=-\epsilon_n(s)$. This leaves 4 (out of the 8) eigenvalues of the lowest Landau levels to be found. Two can easily be identified: $\epsilon_0=-s\Delta_{\text{so}}$ (for $n=0$) and $\epsilon_1=0$ (for $n=1$) both satisfy the quartic equation. The latter implies that a spin-degenerate zero-energy Landau level survives in this context. The two remaining eigenvalues, $\epsilon_-$ and $\epsilon_+$, can be estimated perturbatively, in the limit $\Delta_{\text{so}}, \hbar\omega_c \ll t_1$, as $\epsilon_+ \approx -s\Delta_{\text{so}}(1-\hbar\omega_c/t_1)$ and $\epsilon_- \approx -s\Delta_{\text{so}}\hbar\omega_c/t_1$. Their dependence on $\Delta_{\text{so}}$ and $\hbar\omega_c$ beyond this perturbative regime is shown in the lower panel of Fig.~\ref{Fig3}. This yields the following ordering of eigenvalues: $0 = \epsilon_1 < |\epsilon_-| < |\epsilon_+| < |\epsilon_0| = \Delta_{\text{so}}$.

The corresponding band structure is shown in the top panel of Fig.~\ref{Fig3}. 
Its description in terms of spin-polarized bands is slightly more involved than before, but the Chern numbers can nevertheless be computed and shown to evolve as follows,
\be
\left\lbrace
\begin{array}{l}
{\cal C}_\uparrow=-1 \; , \; {\cal C}_\downarrow=-2 \; , \; \text{for} \; -|\epsilon_0| < E_F < -|\epsilon_+|
\\
{\cal C}_\uparrow=0 \; , \; {\cal C}_\downarrow=-2 \; , \; \text{for} \; -|\epsilon_+| < E_F < -|\epsilon_-|
\\
{\cal C}_\uparrow=1 \; , \; {\cal C}_\downarrow=-2 \; , \; \text{for} \; -|\epsilon_-| < E_F < 0
\\
{\cal C}_\uparrow=2 \; , \; {\cal C}_\downarrow=-1 \; , \; \text{for} \; 0 < E_F < |\epsilon_-|
\\
{\cal C}_\uparrow=2 \; , \; {\cal C}_\downarrow=0 \; , \; \text{for} \; |\epsilon_-| < E_F < |\epsilon_+|
\\
{\cal C}_\uparrow=2 \; , \; {\cal C}_\downarrow=1 \; , \; \text{for} \; |\epsilon_+| < E_F < |\epsilon_0|
\end{array}
\right.
\ee
indicating that a QH phase is preserved at low energy, since the total Chern number never vanishes. This phase is peculiar, however, as it is characterized by edge states with an unbalanced spin population: $|{\cal C}_\uparrow| \neq |{\cal C}_\downarrow|$. For example, two spin-up and a single spin-down counterpropagating state coexist on the same edge for $0 < E_F < |\epsilon_-|$. A given spin species can even become fully gapped, as testified by vanishing spin Chern numbers, giving rise to spin-polarized edge state transport over a tunable and quite large energy window $|\epsilon_+-\epsilon_-| \approx \Delta_{\text{so}}(1-2\hbar\omega_c/t_1)$.

A QSH phase can be generated close to zero energy by lifting the spin-degeneracy of the remaining zero-energy Landau level with an arbitrarily small exchange field, deriving from 
\be
\label{eq:Zee}
H_{\text{ex}} = s\Delta_{\text{ex}}\sigma_0\eta_0 \; ,
\ee
where $\Delta_{\text{ex}}$ quantifies the magnitude of the effect.
The corresponding tight-binding expression is given by
\be
{\cal H}_{\text{ex}} = \Delta_{\text{ex}}\sum_i c_{i,\alpha}^\dagger s_z^{\alpha\beta} c_{i,\beta} \; ,
\ee
using the same notations as before. This yields for the spin Chern numbers at low energy:
\be
{\cal C}_\uparrow=-{\cal C}_\downarrow=1 \; , \; \text{for} \; |E_F| < \text{min}(|\Delta_{\text{ex}}| \; , \; |\epsilon_-|-|\Delta_{\text{ex}}|) \; .
\ee
The total Chern number is zero and, contrary to the case of layer-symmetric spin-orbit coupling, this time the $\mathbb{Z}_2$ invariant $\nu=1$, signaling that the QSH phase is non-trivial. The energy window where this phase can be observed, i.e. the maximum value of the QSH gap, is bounded by the value $|\epsilon_-|/2$, a lower bound of which is given by the perturbative limit $\Delta_{\text{QSH}}^{\text{max}} \geq \Delta_{\text{so}}\hbar\omega_c/(2t_1)$, as can be checked in the lower panel of Fig.~\ref{Fig3}. The QSH gap could thus potentially reach several tens of meV, although, in graphene-based systems, it will effectively be limited by the highest achievable value of spin splitting which should be much smaller\,\footnote{Zeeman splitting can nevertheless be enhanced by tilting the magnetic field with respect to the perpendicular axis, since the Zeeman term is proportional to the total magnetic field $B = B_\parallel + B_\perp$. This avoids the use of large perpendicular magnetic fields which can trigger many-body instabilities.}. In this respect, and despite the need for a perpendicular magnetic field, our proposal offers two advantages with respect to that of Ref.~\cite{Qiao12}, where it was recently shown that gated bilayer graphene could be turned into a $\mathbb{Z}_2$ topological insulator for sufficiently strong Rashba spin-orbit coupling: the strength of spin-orbit coupling need not exceed a critical value, and spin-orbit coupling need not be present in both layers. The latter condition is particularly convenient if one considers that the most promising chance of inducing (intrinsic) spin-orbit coupling in graphene as of today is arguably by depositing adatoms on its surface \cite{Weeks11,Shevtsov12,Jiang12}. 

\subsection{Mechanism II: perpendicular electric field}

Let us now exhibit another mechanism of symmetry-breaking which can provide a loophole to circumvent the intrinsic difficulty of generating a non-trivial QSH phase in bilayer graphene. Forgetting momentarily about spin-orbit coupling, let us go back to the Hamiltonian of Eq.~(\ref{eq:BG}) and consider the effect of an electric field applied perpendicularly to the bilayer,
\be
H_U = U\sigma_0\eta_z \; . 
\ee
This term opens a gap in the energy-momentum dispersion relation by breaking the layer symmetry. It can be implemented in a tight-binding model using the following expression:
\be
{\cal H}_U = -U\sum_{i \in 1}c_i^\dagger c_i + U\sum_{i \in 2}c_i^\dagger c_i \; .
\ee

\begin{figure}[]
\begin{center}
\includegraphics[angle=0,width=1.0\linewidth]{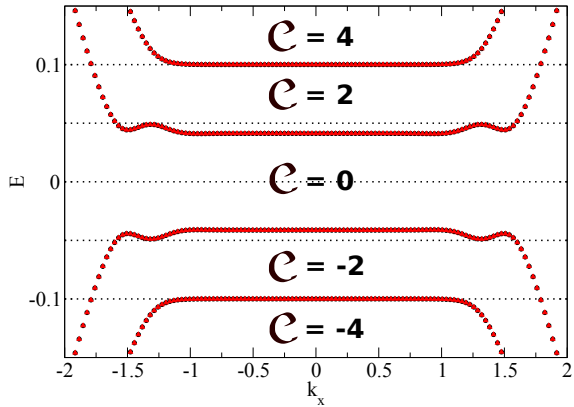}
\includegraphics[angle=0,width=1.0\linewidth]{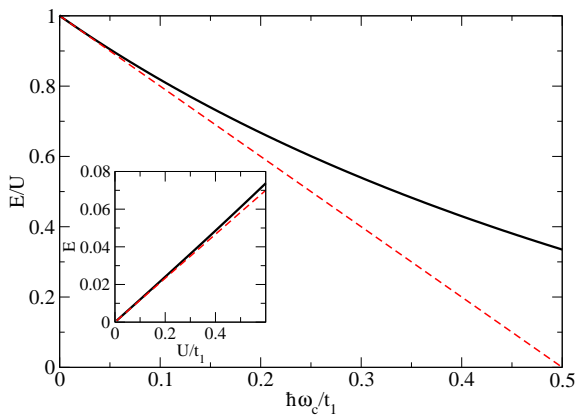}
\caption{(Color online): (Top panel) Effect of an electric field applied perpendicularly to the plane (yielding a layer potential asymmetry $U=0.1$) on the lowest Landau level of bilayer graphene. A gap is opened arising form the lifting of the layer degeneracy for the lowest Landau level. Unspecified parameter values are the same as in Fig.~\ref{Fig2}. (Bottom panel): Dependence of the eigenvalue $\epsilon_-$ on the magnetic field for $U=0.01$. The dashed line is the analytical prediction (valid in the perturbative limit $\hbar\omega_c \ll t_1$) and the thick line is the numerical calculation. Inset: Dependence of $\epsilon_-$ on the perpendicular electric field for $\Delta_B=0.05$.}
\label{Fig4}
\end{center}
\end{figure}

\begin{figure}[]
\begin{center}
\includegraphics[angle=0,width=1.0\linewidth]{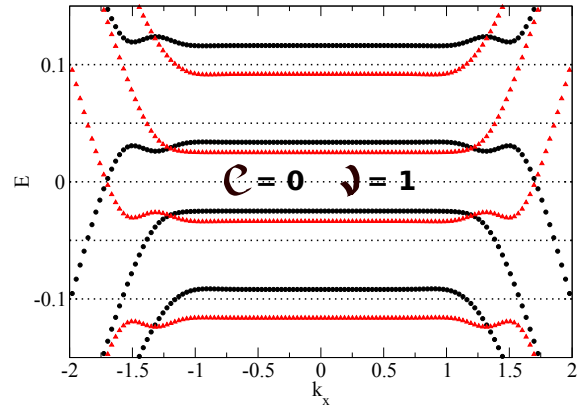}
\caption{(Color online): Effect of an electric field applied perpendicularly to the plane (yielding a layer potential asymmetry $U=0.1$) on the lowest Landau level of bilayer graphene, in the presence of a spin-splitting term $\Delta_{\text{ex}}=0.075$. When this spin splitting exceeds a critical value, a band crossing takes place giving rise to a non-trivial QSH phase. Unspecified parameter values are the same as in Fig.~\ref{Fig2}.}
\label{Fig5}
\end{center}
\end{figure}

The derivation of the Landau level spectrum requires solving once more a quartic equation $\epsilon_n^4 + \alpha_n\epsilon_n^2 + \beta_n\epsilon_n + \gamma_n = 0$, with a non-vanishing linear term $\beta_n\neq0$:
\be
\label{eq:QHU}
\left\lbrace
\begin{array}{l}
\alpha_n = -\left(t_1^2 + 2U^2 + (2n-1)\Delta_B^2\right)
\vspace*{0.2cm}
\\
\beta_n = -2\tau U\Delta_B^2
\vspace*{0.2cm}
\\
\gamma_n = U^2(U^2+t_1^2)  - (2n-1)\Delta_B^2U^2 + n(n-1)\Delta_B^4
\end{array}
\right.
\; .
\ee
Taking into account the spin-degneracy of the spectrum and the additional symmetry $\epsilon_n(-\tau)=-\epsilon_n(\tau)$, one is left with two eigenvalues to compute for the lowest energy Landau levels, one of which can be easily seen to be $\epsilon_+=\tau U$. The other one, $\epsilon_-$, must be computed numerically. In the limit $U, \hbar\omega_c \ll t_1$, it can be estimated perturbatively \cite{McCann06} as $\epsilon_- \approx \tau U(1-2\hbar\omega_c/t_1)$. Its dependence on $U$ and $\hbar\omega_c$ beyond this perturbative regime is shown in the lower panel of Fig.~\ref{Fig4}.
Hence, the QH phase has now been gapped by the perpendicular electric field at low energy (see top panel of Fig.~\ref{Fig4}), yielding the following pattern for the Chern number:
\be
\left\lbrace
\begin{array}{l}
{\cal C}_\uparrow={\cal C}_\downarrow=-1 \; , \; \text{for} \; -|\epsilon_+| < E_F < -|\epsilon_-|
\\
{\cal C}_\uparrow={\cal C}_\downarrow=0 \; , \; \text{for} \; |E_F| < |\epsilon_-|
\\
{\cal C}_\uparrow={\cal C}_\downarrow=1 \; , \; \text{for} \; |\epsilon_-| < E_F < |\epsilon_+|
\end{array}
\right.
\ee
However, the layer degeneracy of the former zero-energy Landau levels has now been lifted, which means that a non-trivial QSH phase can once again be generated at low energy, provided some spin-degeneracy lifting mechanism overcomes the gap $|\epsilon_-|$. This can be achieved either by layer-symmetric spin-orbit coupling (\ref{eq:KM}) or by an exchange term (\ref{eq:Zee}). At a critical value of the spin splitting $|\Delta_{\text{ex}}|=|\epsilon_-|$, the lowest bands will cross and give rise to a QSH phase
\be
\label{eq:QSH2}
{\cal C}_\uparrow=-{\cal C}_\downarrow=1 \; , \; \text{for} \; |E_F| < \text{min}(|\Delta_{\text{ex}}| - |\epsilon_-| \; , \; U-|\Delta_{\text{ex}}|) \; ,
\ee
characterized by a single pair of counter-propagating spin-polarized edge states (see Fig.~\ref{Fig5}). Once again the total Chern number vanishes while the $\mathbb{Z}_2$ invariant $\nu=1$, indicating the non-trivial character of the QSH phase. Provided the critical spin splitting could be achieved, the maximum value of the QSH gap would this time be bounded by the value $(U-|\epsilon_-|)/2$,  an upper bound of which is given by the perturbative limit $\Delta_{\text{QSH}}^{\text{max}} \leq U\hbar\omega_c/t_1$, as can be checked in the lower panel of Fig.~\ref{Fig4}.

As a closing remark, we note that this second mechanism of symmetry-breaking shares with that demonstrated in \cite{Qiao12} the property of having edge states in the low-energy region of Eq.~(\ref{eq:QSH2}) that are not only spin-polarized, but that can also be valley-polarized. This can be traced back to the lifting of valley degeneracy by the perpendicular electric field, as is apparent in the values of the Landau levels given below Eq.~(\ref{eq:QHU}). This valley polarization actually translates into an additional topological protection, encoded in the valley Chern index $\tilde{\nu} = \frac{1}{2}\sum_\tau\tau{\cal C}_\tau$, with ${\cal C}_{\tau}=\sum_s{\cal C}_{\tau,s}$. In the energy region of Eq.~(\ref{eq:QSH2}), this index verifies $\tilde{\nu}=1$, indicating that the low-energy phase is a so-called quantum valley Hall phase. The latter is entirely analogous to a QSH phase, if one exchanges spin and valley indices: it is characterized by valley-polarized counter-propagating edge states, which can thus only be backscattered by short-range (valley-coupling) disorder. Hence, the low-energy phase of Eq.~(\ref{eq:QSH2}) should be immune to spin-mixing perturbations as long as valleys remain uncoupled\,\footnote{In particular, valleys will \textit{de facto} be coupled in armchair-terminated ribbons. The quantum valley Hall phase could however arise in zigzag-terminated ribbons (see \cite{Qiao12}).}.

\subsection{Discussion}

Now that we have identified the regimes in which a non-trivial QSH phase could arise in bilayer graphene, and that we have roughly estimated the order of magnitude of the associated energy gap $\Delta_{\text{QSH}}$, let us conclude this section by making a few comments on the experimental relevance of our results. Until now, we have made the natural assumption of disregarding the effect of disorder in our system, since one of the essential features of a topological phase is its robustness with respect to disorder. The presence of the latter could nevertheless prove problematic if the typical strength of disorder $\delta_{\text{dis}} \gg \Delta_{\text{QSH}}$. The available experimental data in graphene-like systems seem to indicate that low-energy disorder is dominated by charge density fluctuations (electron-hole puddles), but the use of BN substrates has been shown to significantly reduce their magnitude \cite{Dean10,Xue11,Mayorov11}.

The main other threat to the QSH phase lies in the various many-body instabilities which have been predicted to occur in bilayer graphene at the Dirac point due to the finite density of states. This could lead to a spontaneous symmetry breaking of the spin-valley SU(4) symmetry in undoped bilayer graphene, causing the emergence of a yet unidentified gapped phase, typically of the order of a few meV \cite{Feldman09,Velasco12,vanElferen12}. Amusingly, a (many-body driven) QSH phase stands among the list of possible candidates \cite{ZhangF11,Barlas12}.

Estimating the importance of disorder and interaction effects eventually boils down to how big a value of the QSH gap could be achieved. If $\Delta_{\text{QSH}}$ lies in the 10 meV range, then the presence of disorder should be harmless to the QSH phase, while actually reducing the effect of the Coulomb interaction. On the other hand, if $\Delta_{\text{QSH}}$ is rather in the 1 meV range, then chances are great that disorder and/or interactions will wash out the picture we described.

\section{Extensions}
\label{sec:Disc}

Let us now briefly discuss extensions of our model to closely related systems. We start by considering different types of stacking orders in bilayer graphene, and then move on to the case of trilayer graphene.

\subsection{Other stackings}

\begin{figure}[]
\begin{center}
\includegraphics[angle=0,width=1.0\linewidth]{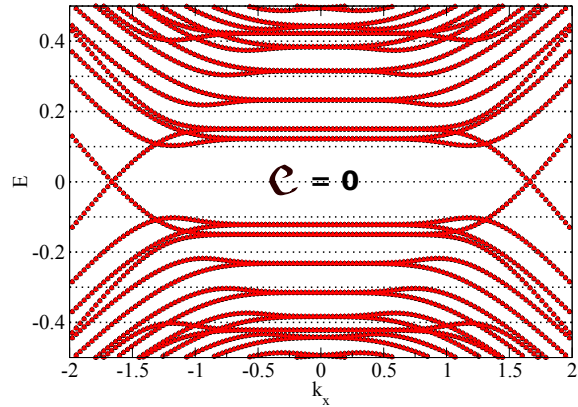}
\caption{(Color online): Landau level spectrum of AA-stacked bilayer graphene with parameter values as in Fig.~\ref{Fig2}. Notice how, at low energy, spin-degenerate counter-propagating edge states lead to a trivial topologically trivial phase.}
\label{Fig6}
\end{center}
\end{figure}

The analysis we performed in this article relied on the assumption of AB (Bernal) stacking for the bilayer. However, other possibilities may occur. One of them is the so-called AA-stacking, where both layers are mirror-symmetric: $A_2$ atoms sit on top of $A_1$ atoms and $B_2$ atoms sit on top of $B_1$ atoms. In this case, following the exact same steps as described in section \ref{sec:Mod}, the Landau level spectrum can be obtained \cite{Hsu10}, $\epsilon_n^{AA}=\pm\sqrt{t_1(t_1 \pm |n|\hbar\omega_c)}$. Contrary to the case of Bernal stacking, the Landau level with lowest energy is no longer necessarily that corresponding to $n=0$, which leads to a peculiar band structure (see Fig.~\ref{Fig6}) where the low-energy physics is described by counter-propagating spin-degenerate edge states, characterized by a trivial ${\cal C}=0$ phase. 
In a sense, the absence of zero-energy Landau levels in this system, which we took as our defining criterium for a 2DDFG, is directly responsible for the absence of a topological order at zero energy. We additionally checked that the symmetry-breaking mechanims investigated in this work are ineffective for the present system.

Besides AB and AA stackings, a whole (continuous) family of bilayers referred to as twisted bilayers can be studied experimentally. Such bilayers are defined by the angle with which the upper layer is twisted from the lower layer. This angle can be probed experimentally by characterizing the induced Moir\'e patterns. Although such systems are also interesting in their own right (and experimentally relevant), they are not well suited to a tight-binding description, especially for small angles, as the low-energy physics requires potentially very long-range hoppings to be taken into account. We will therefore not discuss them any further, and we refer the reader to other approaches developed in the literature to address their properties (see for example \cite{deGail11}).

Likewise, so-called double layer graphene \cite{Ponomarenko11} -- a bilayer where the coupling between the layers is solely capacitive (transverse hopping is zero) -- crucially requires electrostatic screening to be taken into account, and therefore lies beyond the scope of this paper.

\subsection{Trilayer graphene}

\begin{figure}[]
\begin{center}
\includegraphics[angle=0,width=1.0\linewidth]{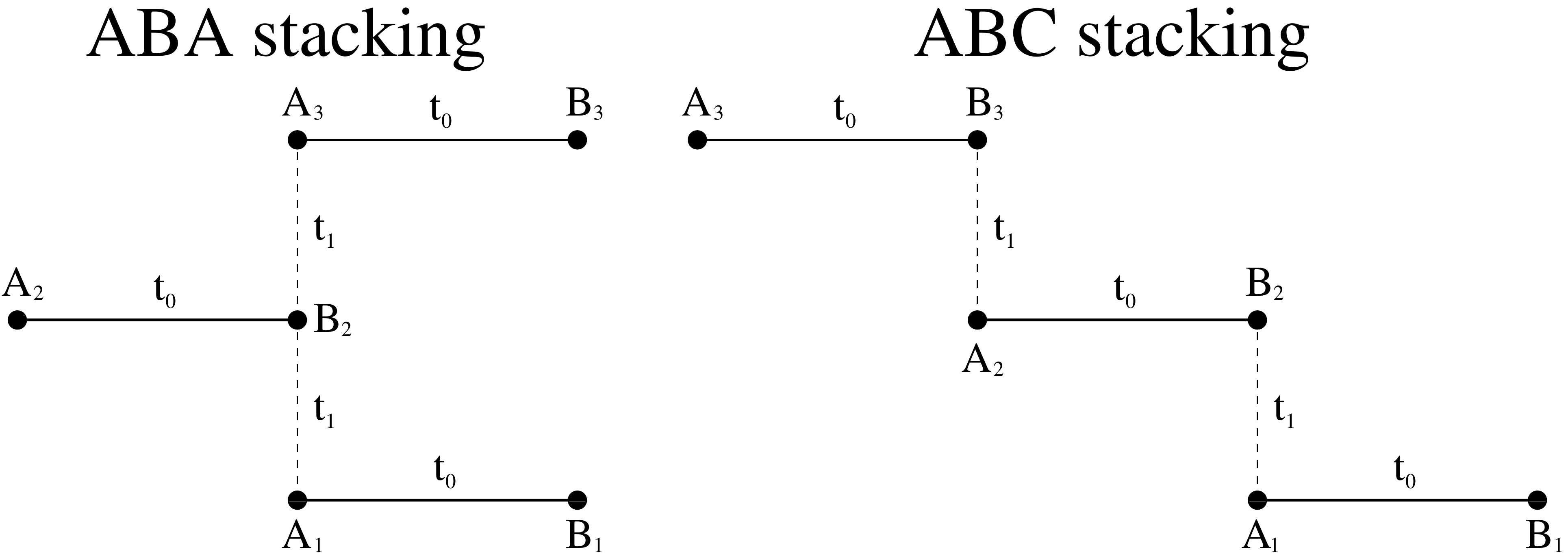}
\caption{Side view of typical stacking sequences of a trilayer of graphite: ABA stacking is mirror-symmetric with respect to the central layer, while ABC stacking can be seen as the natural extension of Bernal stacking in the bilayer (see Fig.~\ref{FigBernal}).}
\label{FigStacking}
\end{center}
\end{figure}

In the light of our understanding of single layer and bilayer graphene, we finish by briefly discussing how much of our previous considerations could find a natural extension in trilayer graphene\,\footnote{Note that, as in bilayer graphene, it has very recently been claimed that a QSH phase could also be induced in gated trilayer graphene in the presence of strong Rashba spin-orbit coupling \cite{Li12}.}. One generally distinguishes two stacking orders (see Fig.~\ref{FigStacking}): ABA stacking, characterized by a combination of linear and quadratic dispersions at low energy, and ABC stacking, characterized by a cubic dispersion and a corresponding diverging density of states at low energy which favors many-body instabilities. 
Regardless of the stacking sequence, the odd number of layers implies that, as in monolayer graphene, including spin-orbit coupling in each layer (via a naive extension of Kane and Mele's model) will yield a non-trivial QSH phase\,\footnote{This is true as long as other possible spin-orbit coupling terms that may arise at low energy in multilayer systems can be neglected \cite{McCann10}.}. The additional presence of a perpendicular magnetic field -- which has experimentally been shown to give rise to a QH effect \cite{Taychatanapat11,Lui11,Bao11,Zhang11} characterized by a spectrum with a 12-fold degenerate zero-energy Landau level and described by the following values for the Chern number,
\be
\left\lbrace
\begin{array}{l}
{\cal C}_\uparrow={\cal C}_\downarrow=-3 \; , \; \text{for} \; -\Delta_{LL} < E_F < 0
\\
{\cal C}_\uparrow={\cal C}_\downarrow=+3 \; , \; \text{for} \; 0 < E_F < \Delta_{LL}
\end{array}
\right.
\; (\Delta_{\text{so}}=0)
\ee
where $\Delta_{LL}$ is the energy of the lowest non-zero Landau level -- will yield a transition from a QH to a non-trivial QSH phase at low-energy (top panel of Fig.~\ref{Fig7}):
\begin{figure}[]
\begin{center}
\includegraphics[angle=0,width=1.0\linewidth]{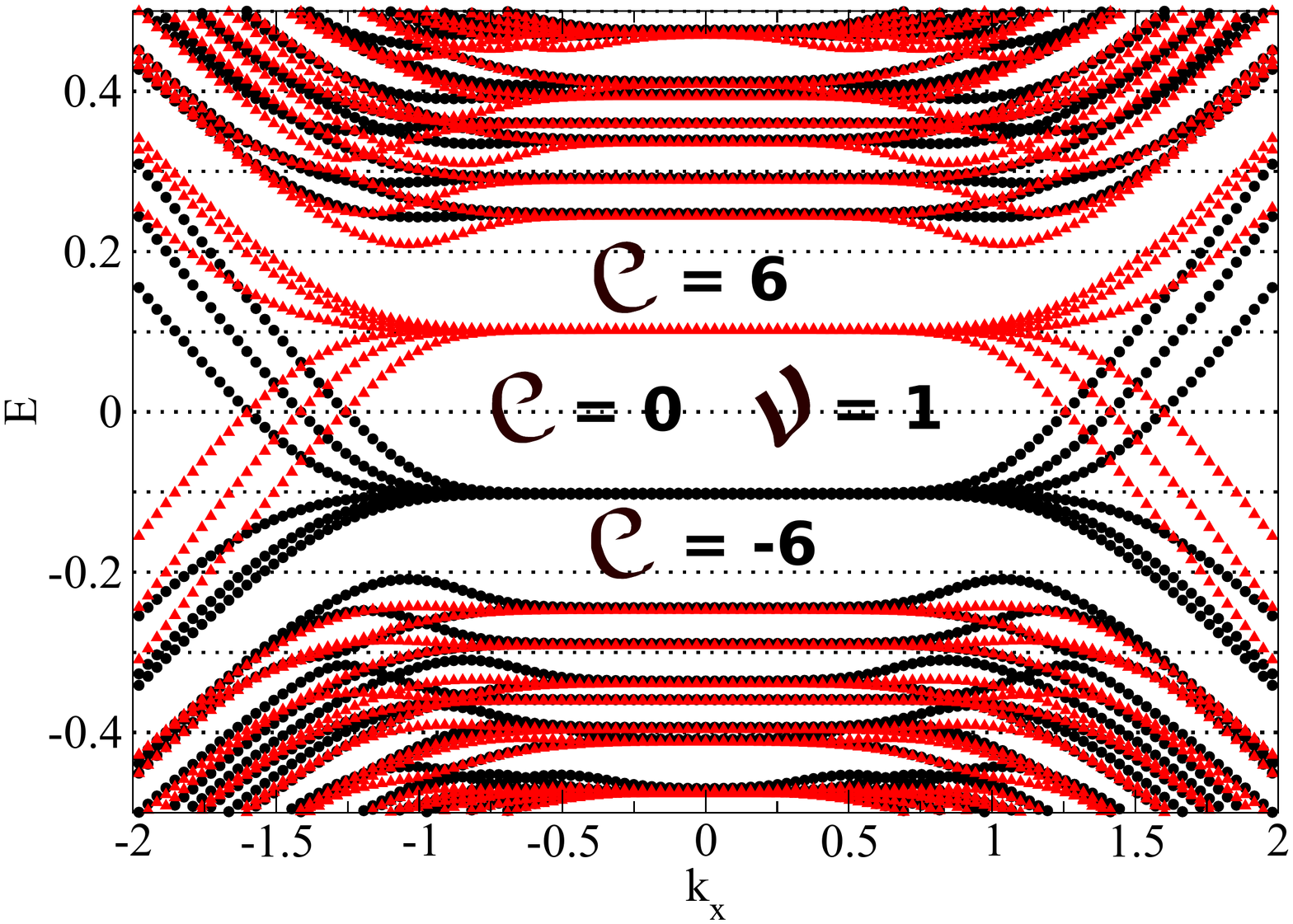}
\includegraphics[angle=0,width=1.0\linewidth]{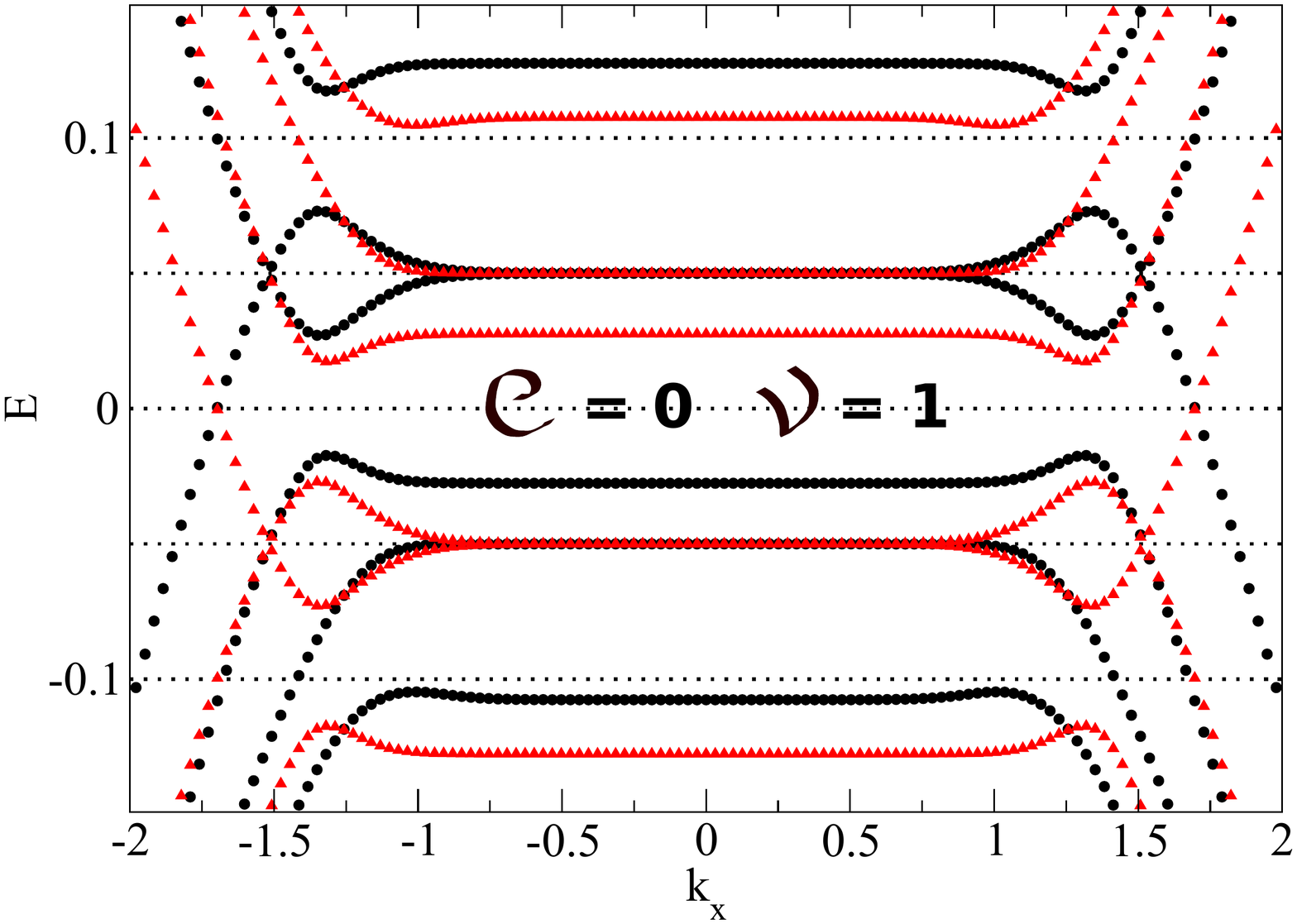}
\caption{(Color online): Landau level spectrum of (ABC-stacked) trilayer graphene with spin-orbit coupling $\lambda_{\text{so}}=0.02$ (top panel), and with both perpendicular electric field ($U=0.1$) and spin-splitting $\Delta_{\text{ex}}=0.05$ (bottom panel). The effect of spin-orbit coupling is completely analogous to that in bilayer graphene (see Fig.~\ref{Fig2}), causing a lifting of spin-degeneracy in the lowest Landau level. This time, however, the odd number of pairs of spin-polarized counter-propagating edge states leads to a non-trivial QSH phase at low energy (top panel). Additionally, and also as in bilayer graphene, the simultaneous presence of a layer-degeneracy lifting electric field and a spin-splitting term can also give rise to a non-trivial QSH phase at low energy, with a single pair of counter-propagating spin-polarized edge states (bottom panel). Once more, unspecified parameter values are the same as in Fig.~\ref{Fig2}.}
\label{Fig7}
\end{center}
\end{figure}
\be
\left\lbrace
\begin{array}{l}
{\cal C}_\uparrow=-{\cal C}_\downarrow=3 \; , \; \text{for} \; |E_F| < \Delta_{\text{so}}
\\
{\cal C}_\uparrow={\cal C}_\downarrow=3 \; , \; \text{for} \; \Delta_{\text{so}} < E_F < \sqrt{\Delta_{LL}^2 + \Delta_{\text{so}}^2}
\end{array}
\right.
\;
\ee
yielding ${\cal C}=0$ and $\nu = 1$ (mod 2) when $|E_F| < \Delta_{\text{so}}$.

\begin{center}
\begin{table*}[hts]
\begin{tabular}{c|c}
\label{table1}
 $N$-layer graphene in the QH regime & Low-energy topological phase \\
\hline\hline
 $N=1$ (Fig.~\ref{Fig1}) & QH with ${\cal C}=\pm2$ \\
\hline
 $N=1$ with $\Delta_{\text{so}}$ (Fig.~\ref{Fig1bot}) & QSH \\
\hline
 $N=2$ (Fig.~\ref{Fig2} top) & QH with ${\cal C}=\pm4$ \\
\hline
 $N=2$ with $\Delta_{\text{so}}$ (Fig.~\ref{Fig2} bottom) & weak QSH with $\nu=0$ (mod 2) \\
\hline
 $N=2$ with $\Delta_{\text{so}}$ only in upper layer (Fig.~\ref{Fig3} top) & QH with ${\cal C}=\pm1$ (spin-unbalanced) \\
\hline
 $N=2$ with $\Delta_{\text{so}}$ only in upper layer, and $\Delta_{\text{ex}}$ & QSH \\
\hline
 $N=2$ with $U$ (Fig.~\ref{Fig4} top) & $\emptyset$ \\
\hline
 $N=2$ with $U$, and $|\Delta_{\text{ex}}|>|\epsilon_-|$ (Fig.~\ref{Fig5}) & QSH + QValleyH \\
\hline
 $N=3$ & QH with ${\cal C}=\pm6$ \\
\hline
 $N=3$ with $\Delta_{\text{so}}$ (Fig.~\ref{Fig7} top) & QSH \\
\hline
\end{tabular}
\caption{Summary of low-energy topological phases in graphene-based 2DDFGs.}
\end{table*}
\end{center}

Exploring further the fate of the Landau level spectrum, we have checked that applying spin-orbit coupling only in the upper layer is (QSH-wise) ineffective. However, applying a perpendicular electric field, through the tight-binding expression
\be
{\cal H}_U = -U\sum_{i \in 1} c_i^\dagger c_i + U\sum_{i \in 3} c_i^\dagger c_i \; ,
\ee
has an interesting effect which distinguishes ABA from ABC stacking. In the latter case, it opens a gap, while in the former it does not (though the QH phase is trivial at low energy, due to counter-propagating states). When an exchange term is taken into account, a QSH phase with only a single pair of counter-propagating spin-polarized states can be accessed (bottom panel of Fig.~\ref{Fig7}). Thus, our second symmetry-breaking mechanism seems to work equally well in trilayer graphene, although its relevance is debatable in the present context since, as mentioned above, a QSH phase could already be obtained in trilayer graphene in the absence of any layer inversion symmetry-breaking. Additionally, the width of the energy window where our mechanism is effective decreases with the number of layers, which can be qualitatively understood as originating from the proliferation of bands (due to the increasing degeneracy of the lowest Landau level).

\section{Conclusion}
\label{sec:Conc}

We have considered different examples of graphene-based 2DDFGs and shown that, in the presence of both spin-orbit coupling and a perpendicular magnetic field, a topological phase transition between a QH and a QSH phase could take place at low energy. An overall summary of the various cases discussed in this article is provided in Table I. While the lifting of spin degeneracy in the Landau level spectrum was the only requirement to observe this transition in monolayer graphene, we showed that a similar prescription proves insufficient in bilayer graphene, yielding a weak QSH phase at low energy. 

We then proceeded to identify several regimes in which a non-trivial QSH phase, characterized by a single pair of counter-propagating spin-polarized edge states, can be induced in bilayer graphene, all of which involved breaking the layer inversion symmetry. We investigated two possible ways of achieving this: (i) by considering the presence of spin-orbit coupling only in the upper layer; (ii) by applying a perpendicular electric field. In both cases, the resulting low-energy phase can then be tuned into a non-trivial QSH phase in the presence of an exchange field: in case (i), an arbitrarily small exchange term suffices, while in case (ii), a non-zero critical value is required. The first of these two cases has the advantage of requiring only small spin splitting (which will, however, effectively control the magnitude of the induced QSH gap) and the presence of spin-orbit coupling only in the upper layer. The latter condition is crucial, as the most promising way to induce sizeable (intrinsic) spin-orbit coupling in graphene is arguably by random adatom deposition \cite{Weeks11,Shevtsov12,Jiang12}. Indeed, the effectively weak intrinsic spin-orbit coupling of carbon remains as of today the main obstacle in the attempt of experimentally detecting the QSH phase in graphene-like systems. In this respect, other recently isolated two-dimensional crystals such as silicene \cite{Lalmi10,Liu11,Ezawa12} or cold atom optical lattices \cite{Mei12} might offer an alternative to probe the physics described in this work. 

\begin{acknowledgments}
This work was supported by STREP ConceptGraphene and EC Contract ERC MesoQMC.
\end{acknowledgments}

\bibliography{BilayerTI}
\bibliographystyle{apsrev}

\end{document}